\documentclass[
preprint,
 amsmath,amssymb,
 aps,
]{revtex4-2}

\usepackage[utf8]{inputenc}
\usepackage{tabularx}
\usepackage{graphicx}
\usepackage{dcolumn}
\usepackage{bm}
\usepackage{amsmath}
\usepackage{amssymb}
\usepackage{subeqnarray}
\usepackage{cases}
\usepackage{enumitem}
\usepackage{xcolor}

\usepackage{esint}

\newcommand{\rmc}{{\rm c}}
\newcommand{\rmi}{{\rm i}}

\newcommand{\rmd}{{\rm d}}
\newcommand{\rmm}{{\rm m}}
\newcommand{\rme}{{\rm e}}

\begin{document}

\title{The electronic and electromagnetic Dirac equations}

\author{Mingjie Li}
\email{ml813@exeter.ac.uk}

\author{ S. A. R. Horsley}
\email{S.Horsley@exeter.ac.uk}
\affiliation{Department of Physics and Astronomy, University of Exeter, Stocker Road, Exeter, UK, EX4 4QL}

\vspace{10pt}

\begin{abstract}
Maxwell's equations and the Dirac equation are the first-order differential relativistic wave equation for electromagnetic waves and electronic waves respectively. Hence, there is a notable similarity between these two wave equations, which has been widely researched since the Dirac equation was proposed.  In this paper, we show that the Maxwell equations can be written in an exact form of the Dirac equation by representing the four Dirac operators with $8\times8$ matrices. Unlike the ordinary $4\times4$ Dirac equation, both spin--1/2 and spin--1 operators can be derived from the $8\times8$ Dirac equation, manifesting that the $8\times8$ Dirac equation is able to describe both electrons and photons. As a result of the restrictions that the electromagnetic wave is a transverse wave, the photon is a spin--1 particle. The four--current in the Maxwell equations and the mass in the electronic Dirac equation also force the electromagnetic field to transform differently to the electronic field. We use this $8\times8$ representation to find that the Zitterbewegung of the photon is actually the oscillatory part of the Poynting vector, often neglected upon time averaging.

\end{abstract}

\maketitle

\vspace{2pc}

\section{Introduction}

Maxwell's equations describe the electromagnetic field, a spin--1 vector field that obeys Bose--Einstein statistics when quantized.  Meanwhile the Dirac equation \cite{DiracEqn}  describes e.g. the electron field, a spin--1/2 field that obeys Fermi--Dirac statistics when quantized.  Despite their very different physical applicability, both equations are first-order differential relativistic wave equations with some surprising parallels. For instance, when the four Maxwell equations are written as a single matrix equation, the result---known as the optical Dirac equation---has an identical structure to the Dirac equation, with spin--1/2 angular momentum operators replacing with their spin--1 counterparts~\cite{OpticalDirac,TopoDirac}.  Other representations have made the same connection via e.g. quaternion algebra~\cite{Li_2020,MaxwellianQM}, and the Riemann-Silberstein vector~\cite{MEsInSpinorNot,ParticleAspectOfEMF,PhotonWave,MajoranaODE}.  This analogy between Maxwell's equations and the  Dirac equation can provide additional insight into some physical problems. 
 For example, we can use the analogy to define the spin operator of the photon, which has been applied to study the spin and orbital angular momentum of the photon~\cite{SOIofLight,feng2022OptDEandSOI,ClassicalED,TransNLongitAM,SpinPhotonics}. The analogy has also been applied to discuss the corresponding Zitterbewegung for the electromagnetic field~\cite{PhotonZitterbewegung,ZitterbewegungOfMasslessParticle}, topological Jackiw--Rebbi states for light~\cite{TopoDirac,EMmodesBoundStatesNTopo,ZeroIndexTopoLight,ZeroIndexMaterial,TopoPhotonics,PhotonicTopoSimulation,DEandOpticalWavePropagation}, surface Maxwell waves~\cite{bliokh2019topological_surface_Maxwell_Waves}, and calculations of thermal equilibrium of the electromagnetic field inside materials~\cite{ClassicalEMatThermalEquilibrium}.

In this paper, we consider a more direct analogy between Maxwell's equations and the Dirac equation.  We find that they can be written in \emph{exactly} the same form, provided we represent the Dirac $\gamma$ matrices as $8\times8$ objects, rather than the usual $4\times4$ matrices.  In this case the Dirac equation is identical to the usual one but with a double counting of the wave degrees of freedom ($8$ complex components rather than $4$), whereas the Maxwell equations emerge as a special case when $2$ of these $8$ components are fixed to be constant. 
  Interestingly---from the conservation of angular momentum---we find that the $8\times8$ Dirac equation automatically contains \emph{two} spin operators, spin--1 and spin--1/2, despite containing a single set of $\gamma$ matrices.  
We explore the transformation properties of this $8$ component wave--function, finding that there are also two possibilities depending on the form of the non-momentum term which is either the four current term (EM field) or the mass term (electron field).  Finally, we explore the Zitterbewegung (`jittering motion') phenomenon predicted by this equation, showing that the same operator equation governs both the photon and electron motion.  For the electromagnetic field it is the familiar oscillating high-frequency part of the Poynting vector that is often neglected upon time averaging.

\section{The Dirac equation in an $8\times8$ matrix representation} \label{DiracEq88Section}

Dirac's famous equation~\cite{DiracEqn} was originally motivated by the desire to have---just as for the Schr\"odinger equation---a first order wave equation in time.  This was achieved as a ``square root'' of the relativistic Klein--Gordon equation~\cite{KleinEqn,GordonEqn}.   Rather than use the non--local operator $\sqrt{\nabla^2}$, Dirac promoted a scalar equation into a matrix one, linear in the momentum $\hat{\bm p}$, through introducing four $4\times4$ matrices $\bm\alpha=(\alpha_x,\alpha_y,\alpha_z)$ and $\beta$,
\begin{equation}\label{OldDiracEq}
	\rmi\hbar\partial_t \Psi=\hat{H}\Psi=\rmc\bm\alpha\cdot\hat{\bm p}\Psi+\beta\rmm\rmc^2\Psi
\end{equation}
with Dirac's conditions being $\alpha_i \alpha_j+\alpha_j\alpha_i=2\delta_{ij}$, $\alpha_i\beta+\beta\alpha_i=0$, and $\beta^2=1$.
There are several different choices of matrices that satisfy these relations, Dirac's original choice being
\begin{equation}
\bm\alpha=\begin{bmatrix}0&\bm\sigma\\ \bm\sigma&0\end{bmatrix}, \qquad \beta=\begin{bmatrix}I_2&0\\0&-I_2\end{bmatrix},\label{eq:dirac-choice}
\end{equation}
with $\bm\sigma=(\sigma_x,\sigma_y,\sigma_z)$ denoting the three Pauli matrices, and $I_2$ the $2\times2$ identity matrix.  As mentioned, the $4\times4$ matrices given above are not the only choices that satisfy Dirac's conditions.  Indeed we could also use higher dimensional objects.  Here we explore $8\times8$ representations of $\bm\alpha$ and $\beta$, constructed though taking a Kronecker product of (\ref{eq:dirac-choice}) with the $2\times2$ identity matrix
\begin{equation}
\bm\alpha=I_2\otimes\begin{bmatrix}0&\bm\sigma\\ \bm\sigma&0\end{bmatrix},
 \qquad
 \beta=I_2\otimes\begin{bmatrix} I_2&0\\0&-I_2\end{bmatrix},\label{eq:88matrices}
\end{equation}
where $\otimes$ denotes the Kronecker product. At this point we have doubled the number of degrees of freedom in the wave--function of the Dirac equation.  One should not worry about these redundant degrees of freedom since the $8\times8$ Dirac equation is just a combination of two ordinary $4\times4$ equations, and the solution is accordingly a combination of two solutions of the two ordinary Dirac equations.

 We now explore another unitarily equivalent representation of the $8\times8$ matrices given above
 \begin{equation}
\boldsymbol{\alpha}'=U\,\boldsymbol{\alpha}\,U^{\dagger}\qquad\beta'=U\,\beta\, U^{\dagger}
 \end{equation}
 where the particular unitary transformation is here given by
 \begin{equation}
 U=\frac1{\sqrt{2}}P_{14}\begin{bmatrix}
M_1 & 0 & M_2 & 0 \\
M_2 & 0 & -M_1 & 0 \\
0 & M_1 & 0 & M_2 \\
0 & M_2 & 0 & -M_1
\end{bmatrix}\label{eq:unitary}
 \end{equation}
 with the $2\times2$ sub--matrices, $M_1={\rm diag}[1,\rmi]$ and $M_2=M_1\sigma_x$ and the $8\times8$ permutation matrix $P_{14}$ of the permutation $1\leftrightarrow4$.  The motivation for the particular transformation (\ref{eq:unitary}) is that the $\bm \alpha$  and $\beta$ matrices can then be written in terms of the matrices corresponding to infinitesimal Lorentz boosts $\bm \kappa$, Euclidean rotations $\bm \theta$, and a diagonal matrix $\eta$: 

\begin{equation}
\bm{\alpha}'=\begin{bmatrix}\bm\kappa&\bm\theta\\ \bm\theta &\bm\kappa\end{bmatrix}.\label{eq:alpha-matrix}
\end{equation}
The component infinitesimal Lorentz boost and rotation matrices in $\bm \alpha'$ are given in the standard form that can be found e.g. in Ref.~\cite{kaku1993}
\begin{equation}
\begin{array}{ccc}
     \kappa_1=\begin{bmatrix}
     0      & -\rmi     & 0     & 0 \\
     \rmi   & 0         & 0     & 0 \\
     0      & 0         & 0     & 0 \\
     0      & 0         & 0     & 0
     \end{bmatrix},
     &
     \kappa_2=\begin{bmatrix}
     0      & 0         & -\rmi     & 0 \\
     0      & 0         & 0         & 0 \\
     \rmi   & 0         & 0         & 0 \\
     0      & 0         & 0         & 0
     \end{bmatrix},
     & 
     \kappa_3=\begin{bmatrix}
     0      & 0         & 0         & -\rmi \\
     0      & 0         & 0         & 0 \\
     0      & 0         & 0         & 0 \\
     \rmi   & 0         & 0         & 0
     \end{bmatrix},
     \\
     \theta_1=\begin{bmatrix}
     0      & 0         & 0     & 0 \\
     0      & 0         & 0     & 0 \\
     0      & 0         & 0     & -\rmi \\
     0      & 0         & \rmi  & 0
     \end{bmatrix},
     &
     \theta_2=\begin{bmatrix}
     0      & 0         & 0         & 0 \\
     0      & 0         & 0         & \rmi \\
     0      & 0         & 0         & 0 \\
     0      & -\rmi     & 0         & 0
     \end{bmatrix},
     & 
     \theta_3=\begin{bmatrix}
     0      & 0         & 0         & 0 \\
     0      & 0         & -\rmi     & 0 \\
     0      & \rmi      & 0         & 0 \\
     0      & 0         & 0         & 0
     \end{bmatrix}.     
\end{array}\label{eq:rotation-boost-matrices}
\end{equation}
although we note that here the Lorentz boosts correspond to using Minkowski's complex $(\rmi \rmc t, x, y, z)$ coordinate system.  The $\beta$ matrix in the Dirac equation is now simply expressed in terms of a diagonal matrix $\eta$ 
$$
\beta'=\begin{bmatrix}-\eta&\\&\eta\end{bmatrix},
$$
where the $\eta$ is given by ${\rm diag}[1,-1,-1,-1]$.

As mentioned above, the $\bm \kappa$ and $\bm\theta$ matrices form the Lorentz group, with the commutation relations between these matrices given by
$$
[\kappa_i,\kappa_j]=\rmi \sum_{k}\epsilon_{ijk}\theta_k,
\qquad
[\theta_i,\theta_j]=\rmi \sum_{k}\epsilon_{ijk}\theta_k,
\qquad
[\kappa_i,\theta_j]=\rmi \sum_{k}\epsilon_{ijk}\kappa_k.
$$
with $\epsilon_{ijk}$ the Levi-Civita symbol.  We now note that the four--dimensional second--rank electromagnetic field tensor $\bm{F }$ can be expressed in terms of the above rotation and boost matrices (\ref{eq:rotation-boost-matrices})
\begin{equation}
\begin{array}{cc}
    \bm F &=-\rmi\bm\kappa\cdot\bm E-\rmi\bm\theta\cdot\bm B,\\
    \bm G &=-\rmi\bm\kappa\cdot\bm B+\rmi\bm\theta\cdot\bm E,
\end{array}\label{eq:field-tensor}
\end{equation}
where $\bm G$ is the dual EM--field tensor (here we use Gaussian units).  We emphasize that the left hand sides of (\ref{eq:field-tensor}) are a $4\times4$ object, and the right hand side consists of a sum over the $4\times4$ matrices (\ref{eq:rotation-boost-matrices}), each weighted by a corresponding electric or magnetic field strength.  This parallel between the Dirac $\bm\alpha$ matrices (\ref{eq:alpha-matrix}) in this new basis, and the electromagnetic field tensor described in terms of rotations and boosts reveals that the $8\times8$ Dirac equation can simultaneously describe both the electron and electromagnetic fields.

To understand how this dual description emerges we first note that, in terms of the field tensor, the Maxwell equations are given by $\partial_{\mu}F^{\mu\nu}=4\pi\rmc^{-1} j^\nu$ and $\partial_{\mu}G^{\mu\nu}=0$. In the notation of (\ref{eq:field-tensor}), the space--time derivative is simply a four--vector operator, $\partial_\mu=(\rmc^{-1}\partial_t,\nabla^T)$ that acts on the left of the equation. Separating the space derivative and the time derivative and using the relations $[0,\nabla^T] (-\rmi\bm\kappa\cdot\bm V)=(\rmi\bm\kappa\cdot\nabla) [0,\bm V^T]^T$ and $[0,\nabla^T](-\rmi\bm\theta\cdot\bm V)=(-\rmi\bm\theta\cdot\nabla)[0,\bm V^T]^T$, then we can re--write Maxwell's equations in the following $8\times8$ form
\begin{equation}
\frac\rmi\rmc \partial_t\begin{bmatrix}0\\ \bm E \\ 0 \\ \rmi\bm B\end{bmatrix}
=\begin{bmatrix}
{\begin{matrix}0&-\nabla\cdot\\ \nabla&\bm0_{3\times3}\end{matrix}} & {\begin{matrix}0&\bm0_{1\times3}\\ \bm0_{3\times1}&\nabla\times\end{matrix}}
\\
{ \begin{matrix}0&\bm0_{1\times3}\\\bm0_{3\times1} &\nabla\times\end{matrix}} &{\begin{matrix}0&-\nabla\cdot\\ \nabla&\bm0_{3\times3}\end{matrix} }
\end{bmatrix}
\begin{bmatrix}0\\ \bm E \\ 0 \\ \rmi\bm B\end{bmatrix}
+\frac{4\pi}\rmc\begin{bmatrix}\rmc\rho\\ -\rmi\bm J \\ \bm0_{4\times1}\end{bmatrix}.\label{eq:maxwell88}
\end{equation}
Noting that the first order differential operator appearing on the right of (\ref{eq:maxwell88}) is $-\rmi\boldsymbol{\alpha}\cdot\nabla$, with $\bm\alpha$ given by (\ref{eq:alpha-matrix})
makes it possible to write the Maxwell's equations in the form of the $8\times8$ Dirac equation introduced above, with the particle mass equal to zero, and the addition of a source term, namely
\begin{equation}\label{PhotonDiracEq}
\rmi\hbar\partial_t\Psi=\rmc\bm\alpha\cdot\hat{\bm p}\Psi+4\pi\hbar\begin{bmatrix}\rmc\rho\\ -\rmi\bm J \\ \bm0_{4\times1}\end{bmatrix},
\end{equation}
where the wave--function is given by
\begin{equation}\label{PhotonWaveFunction}
\Psi=\begin{bmatrix}0\\ \bm E \\ 0 \\ \rmi\bm B\end{bmatrix}.
\end{equation}
The electromagnetic field can thus be described exactly using an $8\times8$ representation of the Dirac equation, with the additional constraint that two components of the wave--function are zero (or at least constant).  As we shall see in the following sections, this constrains the choice of spin operator, as well as the transformation properties of the wave--function, thus allowing a single equation to describe both spin--1/2 and spin--1 particles.

By comparison, the electron wave--function is given by
\begin{equation}\label{ElectronWaveFunction}
\Psi=\begin{bmatrix}\rmi g\\ \bm\varphi\\ f\\ \rmi\bm\chi\end{bmatrix},
\end{equation}
where we have used the notation of~\cite{Li_2020} to define the scalars and vectors $f,g,\bm\varphi,\bm\chi$ that determine the electronic field.  As discussed in~\cite{Li_2020}---where a Maxwell--like description of the electronic field was developed---these variables double count the degrees of freedom in the Dirac equation, although the two sets are uncoupled, as is obvious from the representation (\ref{OldDiracEq}).  However, in performing the unitary transformation (\ref{eq:unitary}), we obscured this tensor product structure and then obtained Maxwell's equations (\ref{eq:maxwell88}). 
 Interestingly, if we apply the inverse unitary transformation to Maxwell's equations, we can return them to the original form (\ref{OldDiracEq}).  However, the constraint that the wave--function takes the form (\ref{PhotonWaveFunction}) prevents us from decomposing this Dirac state into two uncoupled copies.

\section{Lorentz transformation of the $8\times8$ Dirac equation}

The Dirac equation, in general, can be separated into two parts, the part that is linear to the four--momentum $\hat{p}_\mu$ and a remaining part which is independent to the momentum $\hat{p}_\mu$ and hence called the non-momentum term. Mathematically, the Dirac equation is separated into a form as
\begin{equation}\label{GenDiracEq}
\gamma^\mu \hat{p}_\mu\Psi=Y,
\end{equation}
where the four $8\times8$ Dirac gamma matrices in chiral representation are
\begin{equation}\label{DiracGammaMatrices}
   \gamma^0=\begin{bmatrix}&I_2\otimes I_2\\I_2\otimes I_2&\end{bmatrix}, \qquad
    \gamma^i|_{i\ne0}=\begin{bmatrix}&I_2\otimes\sigma_i\\ -I_2\otimes\sigma_i&\end{bmatrix}. 
\end{equation}
Since the Dirac gamma matrices are now $8\times8$ matrices with more degrees of freedom introduced by having the Kronecker product with $I_2$ as shown in (\ref{DiracGammaMatrices}), there are infinite possibilities of the transformation matrix $N$ for the gamma matrices that keeps the left--hand side $\gamma^\mu\hat{p}_\mu$ Lorentz invariant, i.e. , the gamma matrices should satisfies $\gamma^{\mu'}=N\gamma^\mu N^\dagger\Lambda^{\mu'}_\mu$ with $\Lambda^{\mu'}_\mu$ the Lorentz transformation matrix. 

However, the requirement of Lorentz invariance of the non--momentum term $Y$ will limit our choice in choosing the transformation matrix $N$. For the electromagnetic field, the non--momentum term $Y$ is given by the four--current,
$$
Y=-\frac{4\pi\hbar}\rmc\begin{bmatrix}T&\\&T\end{bmatrix}\begin{bmatrix}\rmc\rho\\ -\rmi\bm j \\ -\rmc\rho \\ -\rmi\bm j\end{bmatrix}
, \qquad
T=\begin{bmatrix}1&0&0&-\rmi\\0&-\rmi&1&0\\0&-\rmi&-1&0\\1&0&0&\rmi\end{bmatrix} .
$$
We demand that the four-current must transform as a four--vector. As a result, there left only one choice for the electromagnetic field to transform, namely,
$$
\Psi'=\begin{bmatrix}L^{-1}\otimes L&\\&L\otimes L^{-1}\end{bmatrix}\Psi
,\qquad
L= \sqrt{\frac{\gamma+1}2}+\bm\sigma\cdot\frac{\gamma\bm v}{\rmc\sqrt{2(1+\gamma)}},
$$
with $\gamma=1/\sqrt{1-v^2/\rmc^2}$ in $L$. That is the only one choice for the electromagnetic field to transform in order to keep the Dirac equation Lorentz invariant.

On the other hand, for the electron field, the non--momentum term $Y$ is given by the mass and the field itself,
$$
Y=-\rmm\rmc\Psi.
$$
Unlike the electromagnetic field, in the $8\times8$ representation, the electron field transform as
$$
\Psi'=\begin{bmatrix}N\otimes L&\\& N\otimes L^{-1}\end{bmatrix}\Psi,
$$
where $N$ is an invertible $2\times2$ matrix and it must be the identity matrix $I_2$. Because the $8\times8$ Dirac equation can be viewed as a combination of two independent $4\times4$ ordinary Dirac equation, there is only one way the electron field is allowed to transform in, which should also be a combination of transformations of two independent fields, namely, a combination of such two transformations of $\Psi'_{4\times1}={\rm diag}[L,L^{-1}]\Psi_{4\times1}$, where $\Psi_{4\times1}$ with a subscript denotes the ordinary $4\times1$ Dirac state.

\section{The spin angular momentum operators for the photon and the electron}
In the free space, whether we are describing the electron or the electromagnetic field, the total angular momentum should be conserved.  This angular momentum divides into two parts: spin, $\boldsymbol{S}$ and orbital, $\boldsymbol{L}$.  We now discuss the conservation of the total angular momentum for the $8\times8$ Dirac equation.  The total angular momentum should be constant in time
\begin{equation}
\frac{\rmd }{\rmd t}\langle\hat{\bm L}+\hat{\bm S}\rangle= 0,\label{eq:conservation_angular}
\end{equation}
where $\langle\dots\rangle$ implies an expectation value, and $\hat{\bm L}=\bm r\times\hat{\bm p}$ is the orbital angular momentum operator, and $\hat{\bm S}$ the spin angular momentum operator.  Using the Heisenberg picture we can take the time derivative of the operators in (\ref{eq:conservation_angular}) as e.g. $\rmd\hat{\boldsymbol{S}}/\rmd t=\rmi\hbar^{-1}[\hat{H},\hat{\boldsymbol{S}}]$ with the $8\times8$ Hamiltonian given by (\ref{OldDiracEq}) with the matrices (\ref{eq:88matrices}).  Calculating the commutator between $\hat{H}$ and the $\hat{\boldsymbol{L}}$ operator, (\ref{eq:conservation_angular}) implies that the spin operator must satisfy
\begin{equation}\label{dtSEq}
    \frac{\rmd \hat{\bm S}}{\rmd t}= -\rmc\bm\alpha\times\hat{\bm p}.
\end{equation}
The $\alpha_{i}$ matrices are the same, whether we are describing electrons or photons.  This seems paradoxical given that the electron is a spin--1/2 particle while the photon is spin--1. In our $8\times8$ representation there are in fact two solutions to the operator equation (\ref{dtSEq}).  As expected, these are the spin--1/2 operator
\begin{equation}\label{SpinHalf}
    \bm S_{\frac12}=\frac\hbar2 \begin{bmatrix}\bm\theta&\bm\kappa\\ \bm\kappa&\bm\theta\end{bmatrix}
\end{equation}
and the spin--1 operator
\begin{equation}\label{PhotonSpin}
   \bm S_1=\hbar \begin{bmatrix}\bm\theta&\\&\bm\theta\end{bmatrix} .
\end{equation}

Which of these is relevant is determined by the form of the electromagnetic wave--function (\ref{PhotonWaveFunction}).  For photons, the spin--1/2 operator should be excluded due to the restriction that its first, $f$ and fifth, $g$ wave--function components should always be zero.  Any eigenstate of any observable should satisfy such a restriction since they are physically measurable. Then, when we apply an operator on the photon wave--function, the state we obtain must be a superposition of these eigenstates. The spin--1/2 operator (\ref{SpinHalf}) will give non--constant elements at the first and fifth rows of the photon wave--function, which violates the restriction for the photon wave--function. Therefore, under such a restriction which is due to the requirement of Lorentz invariance, the photon should be a spin--1 particle.

On the other hand, for the electron, the spin--1 operator should be excluded. Because the Dirac equation for the electron can also be represented by $4\times4$ matrices and in this representation, there is only the spin--1/2 operator. The electron is therefore a spin--1/2 particle since the spin of the electron is independent of the representation of the Dirac equation.

\section{The Zitterbewegung of the photon}

The velocity $\rmd \bm r/\rmd t=\langle\rmi\hbar^{-1}[\hat{H},\bm r]\rangle=\rmc\langle\bm\alpha\rangle$  does not commute with the Hamiltonian. As a consequence, there is the Zitterbewegung, recognised for the electronic Dirac equation as the oscillating high--frequency component of the velocity expectation value. To discuss the Zitterbewegung of the electromagnetic field, we write the velocity operator $\bm \alpha$ in Heisenberg picture
\begin{equation}
\bm\alpha(t) = \rme^{\frac\rmi\hbar t \hat{H}}\bm\alpha\rme^{-\frac\rmi\hbar t \hat{H}},\label{eq:alphat}
\end{equation}
where $\bm\alpha$ is the constant matrix defined in the previous sections, and $\hat{H}$ is the $8\times8$ free space Hamiltonian defined in (\ref{OldDiracEq}), with $m=0$.  Integrating the Heisenberg equations of motion as described in~\cite{dirac1957}, we can rewrite the velocity operator as   
\begin{equation}\label{alphaTime}
\bm\alpha(t) =(\bm\alpha-\rmc\hat{\bm p}\hat{H}^{-1})\rme^{-2\frac\rmi\hbar t \hat{H}}+\rmc\hat{\bm p}\hat{H}^{-1}.
\end{equation}

The first term in (\ref{alphaTime}) with the factor of $\exp(-2\frac\rmi\hbar t H)$ indicates an oscillatory motion and is recognized as the Zitterbewegung, here for both the electronic and electromagnetic field.

In the monochromatic case, namely, the particle being in the eigenstate of energy $\varepsilon$, we would not observe the Zitterbewegung due to the fact that $\langle \varepsilon| \bm\alpha \hat{H}+\hat{H}\bm\alpha|\varepsilon\rangle=2\varepsilon\langle\varepsilon|\bm\alpha|\varepsilon\rangle=2\rmc\langle\varepsilon|\hat{\bm p}|\varepsilon\rangle$, which is time independent.  In electromagnetism this calculation is equivalent to finding the time averaged Poynting vector ${\rm Re}[\boldsymbol{E}^{\star}\times\boldsymbol{B}]$.

As we have just established, to exhibit Zitterbewegung, the wave--function must not be an eigenstate of energy.  We now consider the general case where the wave--function is written as a superposition of all energy eigenstates,
\begin{equation}\label{PsiWithMultipleK}
\Psi(t)=\int_k\left[ \Psi(\bm k;+)\rme^{\rmi(\bm k\cdot \bm r-\omega t)}+\Psi(\bm k;-)\rme^{\rmi(\bm k\cdot \bm r+\omega t)}\right]\rmd^3 \bm k ,
\end{equation}
with $\omega=\omega(\bm k)$  the positive solution of the energy--momentum relation, and where we have given separate expressions $\Psi(\boldsymbol{k};\pm)$ for the positive and negative energy expansion coefficients. Substituting the general wave--function (\ref{PsiWithMultipleK}) into the expectation value for the velocity operator (\ref{alphaTime}), we find the Zitterbewegung for the electromagentic field
\begin{multline}\label{ZittPhoton}
\langle\Psi(t=0)|(\bm\alpha-\rmc\hat{\bm p}\hat{H}^{-1})\rme^{-2\frac\rmi\hbar t \hat{H}}|\Psi(t=0)\rangle
\\[10pt]
=\frac{(2\pi)^3}{\varepsilon} \int_{k}\left[ \Psi(\bm k;-)^\dagger \bm\alpha\Psi(\bm k;+) \rme^{-2\rmi\omega  t}+\Psi(\bm k;+)^\dagger \bm\alpha\Psi(\bm k;-) \rme^{2\rmi\omega  t}\right]\rmd^3\bm k
\end{multline}

where we have divided by $\varepsilon$, the total energy of the field, equivalent to normalizing the electromagnetic wave--function.  Given that the Poynting vector is the energy velocity times the energy density, this normalization also makes (\ref{ZittPhoton}) equal to the average electromagnetic energy velocity, in parallel with the equivalent electronic calculation.

As for the electron, the above result (\ref{ZittPhoton}) shows that the Zitterbewegung is a contribution from the positive--energy and the negative--energy components.  This is important for the classical electromagnetic field, as $\bm E$ and $\bm B$ are always real valued since they are both physically measurable. Consequently, the positive--energy and the negative--energy components in (\ref{PsiWithMultipleK}) always come in pairs.  Hence the classical electromagnetic wave function never occurs as a pure eigenstate of energy.

From the above argument we can observe that the Zitterbewegung is the oscillating part of the total classical Poynting vector.  To see this we revert to the Schrodinger picture and write the electromagnetic field $\bm E$ and $\bm B$ in the following form
$$
\bm E=\bm E(\bm r)\rme^{-\rmi\omega t}+\bm E^*(\bm r)\rme^{\rmi\omega t}, \qquad 
\bm B=\bm B(\bm r)\rme^{-\rmi\omega t}+\bm B^*(\bm r)\rme^{\rmi\omega t}.
$$
Substituting these expressions into the photon wave--function (\ref{PhotonWaveFunction}), the integrand of time varying expectation value of $\boldsymbol{\alpha}$ becomes
\begin{equation}\label{EplusEconj}
 \frac12\Psi^\dagger\bm\alpha\Psi=\bm E(\bm r)\times\bm B(\bm r)\rme^{-2\rmi\omega t}+\bm E^*(\bm r)\times\bm B^*(\bm r)\rme^{2\rmi\omega t}+ \bm E^*(\bm r)\times\bm B(\bm r)+ \bm E(\bm r)\times\bm B^*(\bm r) .
\end{equation}
The sum of the first two terms on the right of (\ref{EplusEconj})---often neglected upon time averaging---can therefore be understood as the equivalent of Zitterbewegung for the electromagnetic field.

\section{Conclusions}

In this work we have highlighted the extremely close symmetry between the Dirac equation---usually used to describe high energy electrons---and the classical Maxwell equations.  Although this has been discussed before~\cite{OpticalDirac,PhotonWave,Li_2020}, here we have found a unified $8\times8$ description to represent the Dirac $\boldsymbol{\alpha}$ and $\beta$ matrices. This $8\times8$ Dirac equation has more degrees of freedom than the usual $4\times4$ equation, and is able to simultaneously describe both the electromagnetic and the electronic fields. This single description reveals some interesting parallels and distinctions between these two fields. 

We found that two components of the eight component wave--function must remain zero for us to recover Maxwell's equations from the  $8\times8$ Dirac equation.  Meanwhile, the electron wave--function is unconstrained and simply contains two independent solutions of the ordinary $4\times4$ Dirac equation, as is clear from the tensor product representation used in (\ref{eq:88matrices}).  We found this additional constraint on the electromagnetic field is crucial for both the transformation properties and the choice of spin operator for these two fields.

 Through examining the non--momentum term in the Dirac equation (the mass times the wave function for the electron field and the current for the electromagnetic field), we determined the transformation laws governing the eight component wave--function in these two cases.  We also analysed the spin angular momentum operator derived from the $8\times8$ Dirac equation. We found we can derive both conserved spin--1 and spin--1/2 operators.  The aforementioned constraint on the eight component wave--function implies the expected result that spin of the electromagnetic field must be determined by the spin--1 operator, whereas for the electron field we must use the spin--1/2 operator.

Finally, we discussed the Zitterbewegung predicted by our $8\times8$ Dirac equation, finding that this `jittering motion' is predicted independent of whether the wave--function is for the electronic or electromagnetic field.  Note that our treatment differs from the discussions of electromagnetic Zitterbewegung in Ref.~\cite{ClassicalZitterbewegung,PhotonZitterbewegung,ZitterbewegungOfMasslessParticle}, which did not make use of the unified $8\times8$ description given here.  In Ref.~\cite{ClassicalZitterbewegung}, the Zitterbewegung was treated more abstractly and none of the references give the connection to classical electromagnetism made here.

We found that for both electromagnetism and electrons we simply need a superposition of positive and negative energy solutions to the Dirac equation.  Whereas this produces the expected result for the electronic field, we find that the classical electromagnetic field---which is real valued and thus always in a superposition of positive and negative energy solutions---exhibits this jittering motion through the oscillations of the classical Poynting vector.  These oscillations are well established in classical electromagnetism and usually neglected upon time--averaging.  Nevertheless, we have shown they are the equivalent of Zitterbewegung, arising due to the non--commutation of the Maxwell `Hamiltonian' operator and the velocity operator.

\section*{Acknowledgements}
Mingjie thanks the China Scholarship Council for financial support.  SARH thanks the Royal Society and TATA for financial support (RPG-2016-186).

\section*{Reference}
\bibliographystyle{iopart-num}
\bibliography{Ref.bib}

\end{document}